\documentclass[aip,jcp,reprint,citeautoscript,groupedaddress,noeprint]{revtex4-2}
\usepackage{amsmath}
\usepackage{booktabs}
\usepackage{graphicx}
\usepackage{placeins}
\usepackage{threeparttable}

\usepackage[dvipsnames]{xcolor}
\definecolor{myblue}{rgb}{0,0,1}
\usepackage[breaklinks=true,colorlinks=true,linkcolor=myblue,urlcolor=myblue,citecolor=myblue]{hyperref}

\begin{document}
\title{Resolving Finite-Size Errors in EOM-CCSD Band Gaps of Solids with Interacting-Bath Dynamical Embedding Theory}
\author{Jiachen Li}
\author{Christopher Hillenbrand}
\author{Christian Venturella}
\author{Enzhi Chen}
\author{Tianyu Zhu}
\email{tianyu.zhu@yale.edu}
\affiliation{Department of Chemistry, Yale University, New Haven, CT 06520, USA}

\begin{abstract}
Periodic equation-of-motion coupled-cluster theory with single and double excitations (EOM-CCSD) has shown promise for quantitative calculations of band structures in solids. However, its steep computational scaling has limited calculations to relatively coarse $k$-point meshes, leading to sizable finite-size errors and discrepant estimates of thermodynamic-limit band gaps in recent benchmarks. In this work, we revisit EOM-CCSD band gaps for ten semiconductors and insulators using interacting-bath dynamical embedding theory (ibDET), a systematically improvable Green's function embedding framework that enables dense Brillouin-zone sampling at modest computational cost. By pushing the $k$-point sampling up to $10\times10\times10$, well beyond the system sizes accessible in canonical periodic EOM-CCSD calculations, we significantly reduce finite-size errors and obtain stable thermodynamic-limit extrapolations. We further compare $G_0W_0$@PBE, $G_0W_0$@HF, and EOM-CCSD on an equal footing
using the same numerical settings in PySCF. We find that EOM-CCSD yields a mean absolute error of 0.32 eV relative to experimental band gaps for a test set of ten semiconductors and insulators, lower than that of $G_0W_0$@PBE. For ZnO, EOM-CCSD also accurately describes the Zn $3d$-band binding energy, despite overestimating the band gap. These results demonstrate that ibDET offers a practical route to high-accuracy many-body electronic structure calculations in periodic systems.
\end{abstract}

\maketitle

\section{Introduction}
The accurate prediction of band structures and fundamental band gaps of solids remains a central challenge in computational materials science. 
While density functional theory (DFT)\cite{kohnSelfConsistentEquationsIncluding1965,parrDensityFunctionalTheoryAtoms1989} is the workhorse for periodic electronic structure calculations, 
its well-known band-gap problem due to delocalization errors\cite{cohenInsightsCurrentLimitations2008,williamsCorrectingDelocalizationError2026} necessitates the use of more rigorous many-body approaches. In recent years, 
periodic coupled-cluster (CC) theory has emerged as a promising, systematically improvable framework for computing both ground-state and excited-state properties of solids\cite{mcclainGaussianBasedCoupledClusterTheory2017,gruberApplyingCoupledClusterAnsatz2018,gaoElectronicStructureBulk2020,pulkinFirstprinciplesCoupledCluster2020,dittmerAccurateBandGap2019,wangExcitonsSolidsPeriodic2020,wangAbsorptionSpectraSolids2021,laughonPeriodicCoupledClusterGreens2022,yePeriodicLocalCoupledCluster2024,yeAdsorptionVibrationalSpectroscopy2024,p.carboneCOAdsorptionPt1112024,moermanFiniteSizeEffectsPeriodic2025,neufeldHighlyAccurateElectronic2023,masiosAvertingInfraredCatastrophe2023,mihmShortcutThermodynamicLimit2021,zhangCoupledClusterTheory2019,voCoreBindingEnergies2025}. 
For the quantitative prediction of band gaps, 
ionization-potential and electron-affinity equation-of-motion coupled-cluster theory with single and double excitations (IP/EA-EOM-CCSD)\cite{stantonEquationMotionCoupled1993,stantonAnalyticEnergyDerivatives1994,nooijenEquationMotionCoupled1995}, 
originally developed for molecular charged excitations, 
has attracted particular attention because it directly accesses valence and conduction quasiparticle energies within a unified wavefunction formalism. 
More recently, periodic coupled-cluster Green's function (CCGF) methods have further expanded the reach of CC theory to spectral properties of realistic solids\cite{laughonPeriodicCoupledClusterGreens2022}. 
Compared to many-body perturbation approaches such as $GW$\cite{hedinNewMethodCalculating1965,reiningGWApproximationContent2018,golzeGWCompendiumPractical2019}, 
EOM-CCSD is especially appealing because of its reduced sensitivity to the mean-field starting point and its capability to treat stronger electron correlation.

Despite its theoretical appeal, 
the practical application of EOM-CCSD to extended systems is severely hindered by its steep computational scaling, 
which restricts band gap calculations to small supercells and coarse Brillouin-zone sampling. 
As a result, EOM-CCSD band gaps can exhibit substantial finite-size errors. 
Recent independent benchmark studies have highlighted the challenge of extracting the thermodynamic limit (TDL) from the limited $k$-point meshes accessible in the calculations\cite{voPerformancePeriodicEOMCCSD2024,moermanExploringAccuracyEquationofmotion2025}. 
In earlier studies, 
EOM-CCSD calculations were typically restricted to $\sim4\times4\times4$ $k$-point sampling for simple solids like silicon, 
requiring finite-size extrapolations or $GW$-assisted hybrid strategies\cite{voPerformancePeriodicEOMCCSD2024,moermanExploringAccuracyEquationofmotion2025}. 
The finite-size convergence of IP/EA-EOM-CCSD quasiparticle energies is governed by a leading $N_k^{-1/3}$ term together with potentially important subleading contributions\cite{moermanFiniteSizeEffectsPeriodic2025},
where $N_k$ is the number of $k$-points. 
Consequently, the inferred TDL band gaps can be highly sensitive to the assumed extrapolation form and the specific small-$k$ data points included in the fit, 
leading to noticeable discrepancies among reported benchmark values\cite{voPerformancePeriodicEOMCCSD2024,moermanExploringAccuracyEquationofmotion2025}. 
While substantial progress has been made in understanding finite-size errors for ground-state periodic CC\cite{gruberApplyingCoupledClusterAnsatz2018,liaoCommunicationFiniteSize2016,mihmShortcutThermodynamicLimit2021,mihmHowExchangeEnergy2023,mihmPowerLawsUsed2021,xingInverseVolumeScaling2024,moermanFiniteSizeEffectsPeriodic2025}, 
the mitigation of finite-size errors in periodic EOM-CCSD remains a major bottleneck. 
These discrepancies reflect the difficulty of sampling the Brillouin zone densely enough to reach the asymptotic finite-size scaling regime.

Our central idea is that, 
if EOM-CCSD band gaps can be evaluated on sufficiently dense $k$-meshes, the residual finite-size error should become small enough that the final TDL result is only weakly dependent on the specific extrapolation procedure. 
To achieve this, 
we employ the recently developed interacting-bath dynamical embedding theory (ibDET)\cite{liInteractingBathDynamicalEmbedding2024}, 
a systematically improvable Green's function embedding framework that constructs a bath representation with general two-particle interactions, allowing local and nonlocal correlation effects to be treated on an equal footing. By partitioning the solid into tractable impurity problems while retaining the frequency-dependent entanglement with the environment, 
ibDET bypasses the prohibitive scaling of canonical EOM-CCSD and is therefore particularly well suited to dense Brillouin-zone sampling.
In this work, we design a computational strategy to obtain accurate full-system EOM-CCSD band gaps at each $k$-mesh, 
ranging from $5\times5\times5$ to $8\times8\times8$ for the full test set, and up to $10\times10\times10$ for MgO and Si.
This strategy uses an embedding-size extrapolation scheme that combines $G_0W_0$@HF and EOM-CCSD solvers within the same ibDET framework to obtain accurate EOM-CCSD band gaps at large $k$-meshes. We then extrapolate the resulting $k$-point-dependent EOM-CCSD band gaps to the TDL and apply careful basis-set corrections. 
In addition, to eliminate ambiguities associated with cross-code comparisons, 
we carry out an equal-footing benchmark of $G_0W_0$@PBE, $G_0W_0$@HF, and EOM-CCSD using matched numerical settings in the PySCF software package\cite{sunPySCFPythonbasedSimulations2018,sunRecentDevelopmentsPySCF2020,sunPythonSimulationsChemistry2026}. 
We apply this strategy to ten prototypical solids, 
including LiF, LiCl, BN, BP, C (diamond), Si, SiC, AlP, MgO, and ZnO (zinc blende), 
thereby clarifying prior discrepancies and establishing reliable reference data.

\section{Theory}
We first review the framework of interacting-bath dynamical embedding theory (ibDET).
For a given periodic system,
ibDET starts with a mean-field wavefunction using periodic Gaussian basis set.
The impurity problem is then defined in the intrinsic atomic orbital plus projected atomic orbital (IAO+PAO)\cite{kniziaDensityMatrixEmbedding2013,cuiEfficientImplementationInitio2020} basis,
which consists of orthogonal and atom-centered local orbitals in the real space.
In this work,
we choose local orbitals of all atoms in the unit cell as the impurity,
then gradually expand the embedding space with three sets of bath orbitals to capture electron correlation from short to long range.

First,
employing the idea of density matrix embedding theory (DMET)~\cite{kniziaDensityMatrixEmbedding2012,kniziaDensityMatrixEmbedding2013,cuiEfficientImplementationInitio2020}, bath orbitals $B_\mathrm{DM}$ that capture local charge fluctuations of the impurity are constructed.
By applying the singular value decomposition (SVD) of the mean-field off-diagonal one-particle reduced density matrix (1-RDM) between the impurity and the environment ($\gamma^\mathrm{imp,env}$),
bath orbitals $B_\mathrm{DM}$ exactly reproduce the impurity block of the mean-field 1-RDM
\begin{equation}
    \gamma^{\text{imp,env}} = B_{\text{DM}} \Lambda V^{\dagger}
\end{equation}

Second, 
bath orbitals $B_{\text{GF}}$ that reproduce the mean-field Green's function of the impurity are constructed~\cite{nusspickelEfficientCompressionEnvironment2020}.
To model the charged excitations,
the mean-field one-body Green's function is discretized on a uniform set of real-axis frequency points $\{\omega_n\}$ for the dynamical entanglement between the impurity and the environment.
The occupied and virtual blocks of the retarded one-body mean-field Green's function are defined as
\begin{align}
    (G_0^\text{occ,MO})_{ii} (\omega_n) = & 
    \frac{1}{\omega_n - \epsilon_i + i \eta} \\ 
    (G_0^\text{vir,MO})_{aa} (\omega_n) = & 
    \frac{1}{\omega_n - \epsilon_a + i \eta}
\end{align}
where $\epsilon$ is the mean-field orbital energy, $\eta$ is the broadening parameter used for $B_{\text{GF}}$, and $i$ and $a$ refer to occupied and virtual molecular orbitals.
After rotating occupied and virtual blocks of $G_0^{\text{MO}} (\omega_n)$ into the local orbital space by the transformation matrix $C^{\text{MO,LO}}$,
the occupied and virtual parts of $B_{\text{GF}}$ are obtained from the SVD of the imaginary part of the off-diagonal $G_0^\text{occ,LO} (\omega_n)$ and $G_0^\text{vir,LO} (\omega_n)$
\begin{align}
    \text{Im} [G_0^\text{occ,LO}]^\text{imp,env} (\omega_n) 
    = & \left [ B_\text{GF}^\text{occ} \Lambda V^{\dagger} \right ] (\omega_n) \label{eq:bgf_occ}\\
    \text{Im} [G_0^\text{vir,LO}]^\text{imp,env} (\omega_n) 
    = & \left [ B_\text{GF}^\text{vir} \Lambda V^{\dagger} \right ] (\omega_n) \label{eq:bgf_vir}
\end{align}
where the final $B_{\text{GF}}$ orbitals are assembled as
\begin{equation}
    B_{\text{GF}} = \left [ 
    B_\text{GF}^\text{occ} (\omega_1), B_\text{GF}^\text{vir} (\omega_1), 
    B_\text{GF}^\text{occ} (\omega_2), B_\text{GF}^\text{vir} (\omega_2), 
    \text{...}
    \right ]
\end{equation}
Because of the relation between the 1-RDM and the one-body Green's function at the static limit,
$B_{\text{GF}}$ can be viewed as the dynamical extension to the static $B_{\text{DM}}$\cite{nusspickelEfficientCompressionEnvironment2020}.
Since bath orbitals $B_{\text{GF}}$ from discretizations at different frequency points are not orthogonal,
a projection step is applied to eliminate the redundancy,
which removes the embedding orbitals having minimal overlap with the full-system Hilbert space and then orthogonalizes the embedding space~\cite{liInteractingBathDynamicalEmbedding2024}.
In the projection step,
a cutoff threshold of 0.01 was used in MgO, LiF, LiCl and 0.05 was used for other systems,
which gives around 40 occupied orbitals and 60 virtual orbitals in the $I \bigoplus B_{\text{DM}} \bigoplus B_{\text{GF}}$ space.

To capture the long-range electron correlation beyond the first two sets of bath orbitals,
the cluster-specific natural orbitals\cite{nusspickelSystematicImprovabilityQuantum2022} are utilized to further expand the existing embedding space.
Recently our group developed an improved scheme\cite{venturellaLowScalingManyBodyGreens2026} for constructing the cluster-specific MP2 density matrix,
which shares the idea of pair natural orbitals (PNOs)\cite{riplingerEfficientLinearScaling2013,wangClusterinMoleculeMethodCombined2022,altunExploringAccuracyLimits2023,liangEfficientImplementationRandom2025} and local natural orbitals (LNOs)\cite{rolikGeneralorderLocalCoupledcluster2011,fishmanDevelopmentLocalNatural2026,songRandomPhaseApproximationbased2026} in local correlation approaches.
To select bath orbitals for strongest hybridizations between the embedding space and the environment,
we construct the MP2 amplitudes as
\begin{align}
    t_{i\tilde{j}}^{\tilde{a}\tilde{b}} = & -\frac{(i\tilde{a}|\tilde{j}\tilde{b})}{\epsilon_{\tilde{a}} + \epsilon_{\tilde{b}} - \epsilon_i - \epsilon_{\tilde{j}}}, \label{eq:1pno_t2_1}\\
    t_{\tilde{i}\tilde{j}}^{a\tilde{b}} = & -\frac{(\tilde{i}a|\tilde{j}\tilde{b})}{\epsilon_a + \epsilon_{\tilde{b}} - \epsilon_{\tilde{i}} - \epsilon_{\tilde{j}}} \label{eq:1pno_t2_2}
\end{align}
In Eq.~\ref{eq:1pno_t2_1} and Eq.~\ref{eq:1pno_t2_2},
the tilde sign over an index denotes an orbital in the embedding cluster $I \bigoplus B_\mathrm{DM} \bigoplus B_\mathrm{GF}$, 
whereas an index without a tilde sign denotes an orbital in the environment. This method was referred to as the ``1-PNO'' scheme in Ref.~\citenum{venturellaLowScalingManyBodyGreens2026}.
The cluster-specific MP2 density matrix is then constructed as
\begin{align}
    \gamma_{ij} = & 2 \delta_{ij} - 2 \sum_{\tilde{k}\tilde{a}\tilde{b}} t^{\tilde{a}\tilde{b}}_{i\tilde{k}} \left [ 2 t^{\tilde{a}\tilde{b}}_{j\tilde{k}} - t^{\tilde{b}\tilde{a}}_{j\tilde{k}} \right ] \\
    \gamma_{ab} = & 2 \sum_{\tilde{i}\tilde{j}\tilde{c}} t^{a\tilde{c}}_{\tilde{i}\tilde{j}} \left [ 2 t^{b\tilde{c}}_{\tilde{i}\tilde{j}} - t^{\tilde{c}b}_{\tilde{i}\tilde{j}} \right ]
    \label{eq:mp2dm_2}
\end{align}
We then derive the cluster-specific natural orbital bath, \(B_{\mathrm{NO}}\), by diagonalizing these MP2 density matrices and retaining the resulting natural orbitals whose fractional occupations exceed a pre-defined threshold. Our final embedding space is thus $I \bigoplus B_\mathrm{DM} \bigoplus B_\mathrm{GF} \bigoplus B_\mathrm{NO}$, where redundant bath degrees of freedom are removed in a final projection step.
This one-environment-index scheme significantly reduces the computational cost to evaluate the MP2 density matrix. For example, for silicon on a \(10\times10\times10\) \(k\)-point mesh, it enables the construction of MP2 natural orbitals for an embedded cluster containing 42 occupied orbitals coupled to an environment of 21,932 virtual orbitals in the corresponding supercell, as described in Eq.~\ref{eq:1pno_t2_2} and Eq.~\ref{eq:mp2dm_2}. 

After constructing the embedding problem,
the impurity solver is used to obtain the self-energy in the embedding space.
In this work,
coupled-cluster Green's function at the EOM-CCSD level (CCGF)\cite{nooijenCoupledClusterApproach1992,nooijenCoupledClusterGreens1993,pengGreensFunctionCoupledCluster2018,zhuCoupledclusterImpuritySolvers2019,sheeCoupledClusterImpurity2019,laughonPeriodicCoupledClusterGreens2022} and space-time $G_0W_0$\cite{liuCubicScaling$GW$2016,wilhelmGWCalculationsThousands2018,wilhelmLowScalingGWBenchmark2021} solvers are utilized.
Then the self-energy in the embedding space is rotated back to the full space,
\begin{equation}
    \Sigma^\text{full}(\omega) =
    R \Sigma^\text{emb}(\omega) R^{\dagger},
    \label{eq:sigmarotation}
\end{equation}
where $R$ is the rotation matrix from the full IAO+PAO space to the embedding space:
\[
R =
\left[
\begin{array}{c|ccc}
I & & & \\
\hline
\rule[-3.5em]{0pt}{8em}
& B_{\mathrm{DM}} & B_{\mathrm{GF}} & B_{\mathrm{NO}}
\end{array}
\right].
\]
With the full-space self-energy that encodes both short- and long-range electron correlations,
the many-body Green's function $G$ can be calculated via the Dyson equation
\begin{equation}
     G^{-1}(\omega) = G_0^{-1}(\omega) - \Sigma^\text{full}(\omega),
\end{equation}
where $G_0$ is the full-space mean-field Green's function.
From the Green's function,
the spectral function can be obtained by
\begin{align}
    A(k, \omega) \equiv & -\frac{1}{\pi} \text{Im} \left[G(k, \omega)\right],
\end{align}
from which the quasiparticle band structure is obtained.

\section{Computational Details}
We studied band gaps of ten semiconductors and insulators including LiF, LiCl, BN, BP, C (diamond), MgO, AlP, Si, SiC and ZnO (zinc blende),
where lattice constants, band gap positions and reference values are taken from 
Refs.\citenum{kimStructuralReconstructionHexagonal2003,miglioPredominanceNonadiabaticEffects2020,zhuAllElectronGaussianBasedG0W02021,voPerformancePeriodicEOMCCSD2024} and given in the Supporting Information (SI).
All ground-state Hartree-Fock (HF) and PBE\cite{perdewGeneralizedGradientApproximation1996} calculations in periodic Gaussian basis sets were carried out using the PySCF quantum chemistry software package~\cite{sunPySCFPythonbasedSimulations2018,sunRecentDevelopmentsPySCF2020,sunPythonSimulationsChemistry2026} 
with Gaussian density fitting. 
The cc-pVDZ-PP basis set and the corresponding effective core potential\cite{petersonSystematicallyConvergentBasis2005} were used for Zn,
and the GTH-HF-rev pseudopotential\cite{hartwigsenRelativisticSeparableDualspace1998} and the GTH-cc-pVDZ basis set~\cite{yeCorrelationConsistentGaussianBasis2022} were used for all other elements.
One-shot $G_0W_0$@HF and $G_0W_0$@PBE calculations were performed with the fcDMFT package\cite{zhuEfficientFormulationInitio2020,zhuAllElectronGaussianBasedG0W02021,zhuInitioFullCell2021,liRestoringTranslationalSymmetry2024}.
Finite-size Coulomb divergence corrections were not used in HF or $G_0W_0$ calculations.

To obtain $G_0W_0$ results in the TDL,
the cubic extrapolation formula proposed in Ref.\citenum{moermanFiniteSizeEffectsPeriodic2025} was used
\begin{equation}\label{eq:fit_cubic}
    E_{N_k} = E_\text{TDL} + A N_k^{-\frac{1}{3}} + B N_k^{-\frac{2}{3}} + C N_k^{-1}
\end{equation}
where $E$ is the band gap and $N_k$ is the number of $k$-points.
For BN, BP, C, AlP, Si and SiC that have an indirect band gap,
Eq.~\ref{eq:fit_cubic} is first fitted with $\Gamma\to\Gamma$ gaps.
To obtain indirect $G_0W_0$ band gaps in the TDL,
we first extrapolated the $\Gamma\to\Gamma$ gaps from $3\times3\times3$ to $8\times8\times8$ to the TDL.
Then the conduction-band offset between the conduction band minimum (CBM) and lowest unoccupied quasiparticle energy at $\Gamma$ point evaluated on the $8\times8\times8$ $k$-mesh was added.
The $G_0W_0$ band gaps in the TDL were fitted with $k$-meshes from $3\times3\times3$ to $8\times8\times8$.
As shown in the SI,
this provides close results to $G_0W_0$ band gaps fitted with $k$-meshes from $3\times3\times3$ to $10\times10\times10$ for MgO and Si. For comparison, we also examine a linear extrapolation form that includes only the leading-order \(N_k^{-1/3}\) finite-size correction~\cite{zhuAllElectronGaussianBasedG0W02021,voPerformancePeriodicEOMCCSD2024}:
\begin{equation}\label{eq:fit_linear}
    E_{N_k} = E_{\mathrm{TDL}} + A N_k^{-1/3}.
\end{equation}
The comparison of $G_0W_0$ band gaps in the TDL using different extrapolation formulas can be found in the SI.

In ibDET calculations,
$3s3p3d4s$ orbitals of Zn and the GTH-SZV basis set for all other elements were used as the pre-defined minimal atomic orbital (MINAO) basis to construct IAOs.
IAO+PAO local orbitals of all atoms in the unit cell were selected as the impurity.
In the frequency-dependent bath orbital ($B_{\text{GF}}$) construction,
uniform real-frequency grids ranging from 0.3 a.u. below the HF valence band maximum (VBM) energy to 0.3 a.u. above the HF conduction band minimum (CBM) energy with a spacing of 0.03 a.u. were used.
In the natural bath orbital ($B_\mathrm{NO}$) construction,
15 canonical virtual orbitals were added for all calculations.
We use ``HF+$GW$'' to denote ibDET calculations in which HF is used as the low-level theory and $G_0W_0$@HF as the impurity solver, and ``HF+CC'' to denote the corresponding ibDET calculations with CCGF as the impurity solver.

To control the two dominant finite-size effects separately, we employ a two-stage extrapolation strategy: we first remove the finite-embedding-size error at each fixed \(k\)-mesh by extrapolating the ibDET results to the full-system limit, and then remove the finite-\(k\)-mesh error by extrapolating the resulting full-system-limit EOM-CCSD gaps to the TDL. In the first stage, to obtain full-system-limit EOM-CCSD results for a given \(k\)-mesh, 
we employed an embedded band-gap extrapolation strategy, in which the ibDET HF+CC band gaps are extrapolated to zero embedding error using the corresponding ibDET HF+\(GW\) embedding error as the extrapolation coordinate
\begin{equation}\label{eq:fit_emb}
    E^\text{HF+CC}_\text{emb} = E^\text{EOM-CCSD}_\text{full} + A ( E^{\text{HF+}GW}_\text{emb} - E^{GW}_\text{full} )
\end{equation}
where the subscript ``emb'' stands for embedding calculations.
Embedding spaces consisting of 260, 280, 300, 320, 340, and 360 orbitals are used in the extrapolation formula in Eq. \ref{eq:fit_emb}.
The number of virtual embedding orbitals for each space is selected as $N^\text{emb}_\text{vir}=3 N^\text{emb}_\text{occ}$.

In the second stage, to obtain EOM-CCSD band gaps in the TDL, we use two strategies for finite-\(k\)-mesh extrapolation.
In the first, ``\(GW\)-assisted'' strategy, following Ref.~\citenum{moermanExploringAccuracyEquationofmotion2025}, the EOM-CCSD band gaps obtained from Eq.~\ref{eq:fit_emb} at different \(k\)-meshes are extrapolated with respect to the finite-size error of \(G_0W_0\)@HF
\begin{equation}\label{eq:fit_cc}
    E^\text{EOM-CCSD}_{N_k} = E^\text{EOM-CCSD}_\text{TDL} + A ( E^{GW@\text{HF}}_{N_k} - E^{GW@\text{HF}}_\text{TDL} )
\end{equation}
We used $k$-mesh from $5\times5\times5$ to $8\times8\times8$ in Eq.~\ref{eq:fit_cc},
which provides similar results to those fitted with $k$-meshes from $5\times5\times5$ to $10\times10\times10$ for MgO and Si, as shown in the SI. In the second, ``linear'' extrapolation strategy, following Ref.~\citenum{voPerformancePeriodicEOMCCSD2024}, we extrapolate the EOM-CCSD band gaps at different \(k\)-meshes directly using Eq.~\ref{eq:fit_linear}.

After obtaining the band gaps at the \(G_0W_0\) and EOM-CCSD levels with double-zeta basis sets, we applied basis-set corrections to the quadruple-zeta level, estimated as the difference between quadruple-zeta and double-zeta band gaps at a finite \(k\)-mesh:
\begin{equation}
    E^\text{M,QZ}_\text{TDL} = E^\text{M,DZ}_\text{TDL} + 
    (E^\text{M,QZ}_{N_k} - E^\text{M,DZ}_{N_k})
\end{equation}
where $\mathrm{M}=G_0W_0@\mathrm{PBE}$, $G_0W_0@\mathrm{HF}$, or EOM-CCSD.
The quadruple-zeta calculations used cc-pVQZ-pp basis set for Zn and GTH-cc-pVQZ basis set for all other elements. 
The corrections were evaluated on a \(6\times6\times6\) \(k\)-mesh for full \(G_0W_0\) and on a \(2\times2\times2\) \(k\)-mesh for full periodic EOM-CCSD, and were then added to the corresponding double-zeta TDL results. \\

\onecolumngrid

\begin{figure*}[hbt]
\centering
\includegraphics[width=0.99\textwidth]{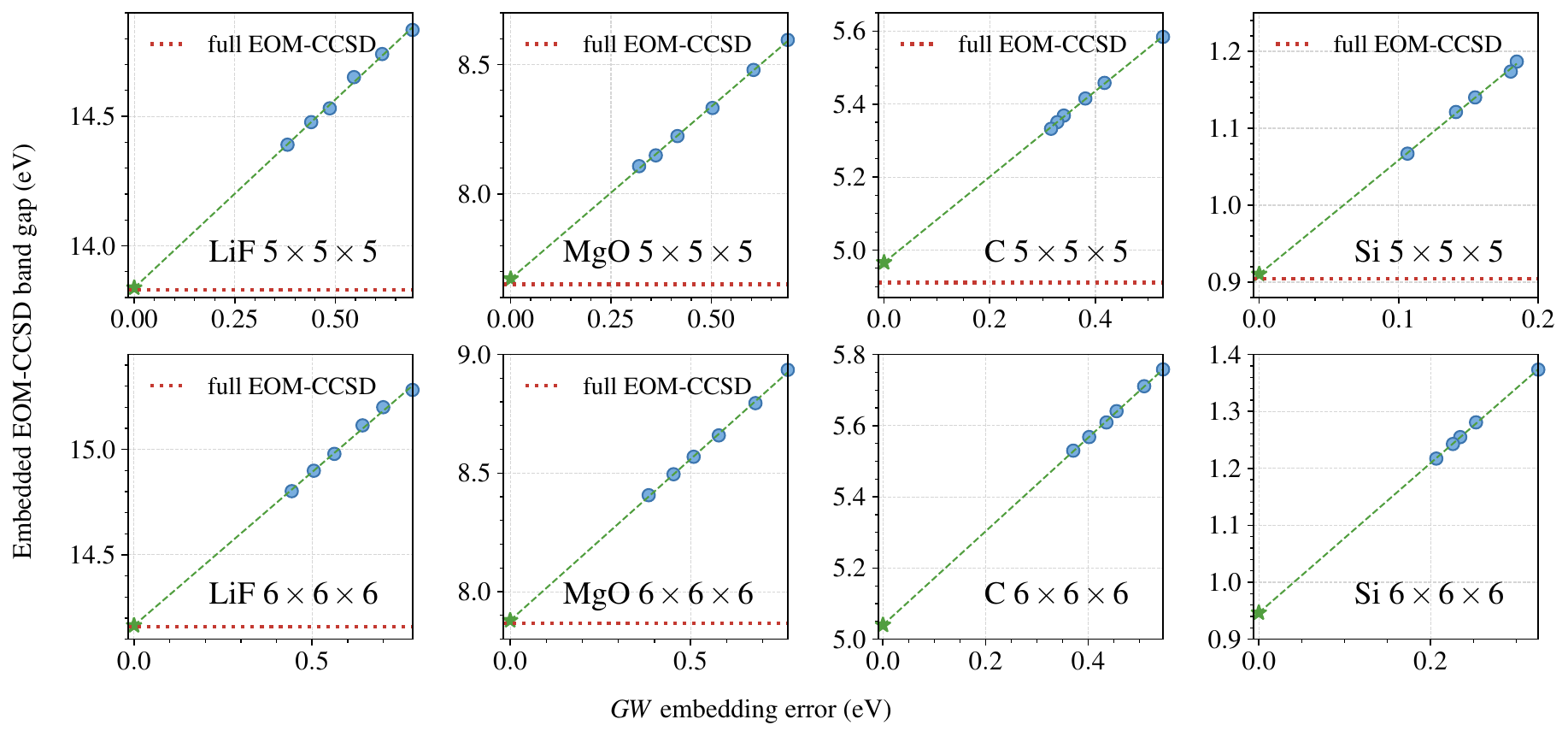}
\caption{Extrapolations of embedded HF+CC band gaps against $GW$ embedding errors to the full-system EOM-CCSD limit at fixed $k$-meshes. The $GW$ embedding error is defined as the difference between HF+GW band gap 
and the full-system $G_0W_0$@HF band gap.
Extrapolated values are indicated by the star symbols.
Full-system periodic EOM-CCSD references obtained from Ref.\citenum{li2026reaching} are shown as horizontal red dashed lines.
GTH-cc-pVDZ basis set and GTH-HF-rev pseudopotential were used.}
\label{fig:emb_error}
\end{figure*}

\twocolumngrid

\section{Results}
\subsection{Convergence of ibDET band gaps to the full-system limit}
We first examine the convergence of ibDET results to the full-system limit at a fixed $k$-mesh. The goal is to assess whether our embedded band-gap extrapolation strategy yields accurate full-system EOM-CCSD band gaps at each $k$-mesh. 
Extrapolations of HF+CC band gaps of LiF, MgO, C, and Si using $5\times5\times5$ and $6\times6\times6$ $k$-meshes against $G_0W_0$@HF embedding errors are shown in Fig.~\ref{fig:emb_error},
where the $G_0W_0$@HF embedding error is defined as the difference between the embedded HF+$GW$ band gap and the full-system $G_0W_0$@HF band gap at a given $k$-mesh. We note that, for each data point in Fig.~\ref{fig:emb_error}, the HF+CC and HF+GW calculations were performed using the same embedding Hamiltonian.
Very recently, Li \textit{et al.}~\cite{li2026reaching} developed a distributed-memory implementation of periodic CCSD that
enabled canonical EOM-CCSD calculations on $k$-point meshes as dense as $6\times6\times6$. 
The full-system periodic EOM-CCSD reference values taken from Ref.\citenum{li2026reaching} are used to judge the quality of embedding size extrapolation in this work.
At each $k$-mesh,
ibDET band gaps with six embedding spaces consisting of 260 to 360 orbitals are used in the extrapolation. 

Fig.~\ref{fig:emb_error} shows that, for different systems and \(k\)-point meshes, the ibDET HF+CC band gaps vary nearly linearly with the corresponding \(G_0W_0\)@HF embedding errors. This behavior supports the embedded extrapolation scheme used to reach the full-system EOM-CCSD limit.
Similar convergence behaviors in all tested systems with $5\times5\times5$ to $8\times8\times8$ $k$-meshes can be found in the SI.
Furthermore,
for medium $k$-meshes where the full-system EOM-CCSD results are available,
the extrapolated HF+CC band gaps agree well with the full-space results.
For LiF and MgO that have a direct band gap,
the errors of extrapolated HF+CC values compared with full-system EOM-CCSD results at $5\times5\times5$ and $6\times6\times6$ $k$-meshes are less than 0.02 eV.
For C and Si that have an indirect band gap,
the errors at $5\times5\times5$ $k$-mesh are around 0.05 eV and 0.01 eV. 
We thus establish the scheme to obtain converged EOM-CCSD band gaps at each $k$-mesh from ibDET calculations.
We also note that, for the largest embedding space consisting of 90 occupied orbitals and 270 virtual orbitals,
the computational cost of the CCGF solver step is about 1800 CPU hours,
which is significantly smaller than full-system EOM-CCSD calculations at medium and large $k$-meshes. \\

\onecolumngrid

\begin{figure*}[hbt!]
\centering
\includegraphics[width=0.99\textwidth]{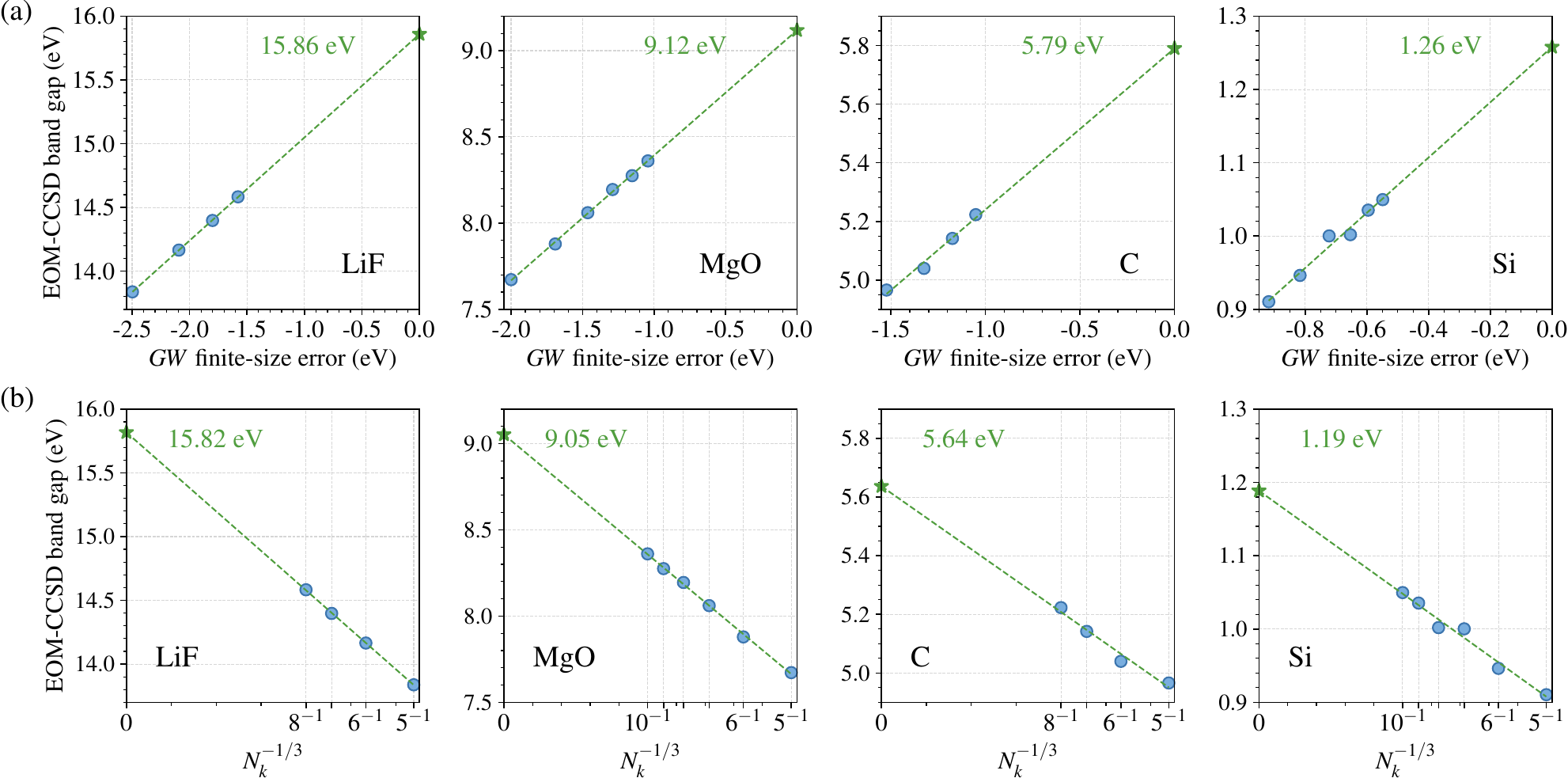}
\caption{
Thermodynamic-limit extrapolations of EOM-CCSD band gaps for LiF, MgO, C, and Si obtained from ibDET calculations. 
(a) \(GW\)-assisted extrapolation, where the EOM-CCSD band gaps are extrapolated against the \(G_0W_0\)@HF finite-size error.
(b) Direct linear extrapolation with respect to \(N_k^{-1/3}\).
The extrapolated TDL values are indicated by star symbols.
The GTH-cc-pVDZ basis set and GTH-HF-rev pseudopotentials were used.
}
\label{fig:cc_tdl}
\end{figure*}

\twocolumngrid

\subsection{Finite \texorpdfstring{$k$}{k}-mesh extrapolation to the thermodynamic limit}
Having obtained full-system-limit EOM-CCSD band gaps at each finite \(k\)-mesh from the ibDET embedding size extrapolation, we next extrapolate these gaps to the TDL. 
Here, we compare two finite-\(k\)-mesh extrapolation schemes previously used in periodic EOM-CCSD benchmark studies. 
The first is the \(GW\)-assisted extrapolation scheme of Ref.~\citenum{moermanExploringAccuracyEquationofmotion2025}, which exploits the similar finite-size behavior of \(G_0W_0\)@HF and EOM-CCSD quasiparticle gaps. 
In this approach, the EOM-CCSD band gaps are extrapolated against the \(G_0W_0\)@HF finite-size error, defined as the difference between the \(G_0W_0\)@HF band gap at a given \(k\)-mesh and its TDL value. 
The second is the direct linear extrapolation scheme of Ref.~\citenum{voPerformancePeriodicEOMCCSD2024}, in which the EOM-CCSD finite-size error is assumed to scale linearly with \(N_k^{-1/3}\).

Fig.~\ref{fig:cc_tdl} compares these two extrapolation schemes for LiF, MgO, C, and Si. 
For LiF and C, we use \(k\)-meshes from \(5\times5\times5\) to \(8\times8\times8\), while for MgO and Si we use \(k\)-meshes from \(5\times5\times5\) to \(10\times10\times10\) to assess the reliability of extrapolations based on smaller $k$-meshes than $10\times10\times10$.
The finite-\(k\)-mesh corrections remain sizable over this range, especially for the wide-gap materials LiF and MgO, indicating that explicit TDL extrapolation is still necessary even with relatively dense \(k\)-point sampling. 
Nevertheless, the data exhibit nearly linear behavior in both extrapolation coordinates once these dense \(k\)-meshes are used. 
The \(GW\)-assisted and linear extrapolations give TDL gaps of 15.86 and 15.82 eV for LiF, 9.12 and 9.05 eV for MgO, 5.79 and 5.64 eV for C, and 1.26 and 1.19 eV for Si, respectively. 
Thus, the two extrapolation schemes differ by only \(0.04\)--\(0.15\) eV for the systems shown. 
The largest difference occurs for diamond, and a smaller but visible difference is also observed for Si; both are indirect-gap systems, for which the finite-\(k\)-mesh determination of the CBM introduces an additional source of uncertainty. 
As shown in the SI, including smaller \(k\)-meshes leads to larger deviations from linearity in both extrapolation schemes.

\begin{table}[hbt!]
\setlength\tabcolsep{5pt}
\caption{EOM-CCSD band gaps of ten semiconductors and insulators in the TDL.
Results from this work were obtained by extrapolating ibDET-derived full-system-limit EOM-CCSD gaps using \(k\)-meshes from \(5\times5\times5\) to \(8\times8\times8\), with both \(GW\)-assisted and direct linear finite-size extrapolation schemes. 
Basis-set corrections to the quadruple-zeta level evaluated at $2\times2\times2$ $k$-mesh were added to the double-zeta EOM-CCSD band gaps. 
Literature values are taken from Refs.~\citenum{voPerformancePeriodicEOMCCSD2024}, \citenum{moermanExploringAccuracyEquationofmotion2025}, and \citenum{li2026reaching}.
All values are in eV.}
\begin{tabular}{cccccc}
\toprule
System & \multicolumn{2}{c}{This work}
& Ref.\citenum{voPerformancePeriodicEOMCCSD2024} 
& Ref.\citenum{moermanExploringAccuracyEquationofmotion2025}
& Ref.\citenum{li2026reaching} \\
\cmidrule(lr){2-3}
& $GW$-assisted & linear & & \\
\midrule
LiF  & 15.92 & 15.88  & 15.43 & 16.19 & 15.93 \\
LiCl & 10.11 & 10.06  & 9.43  & 9.90  & 10.18 \\
BN   & 6.68  & 6.55   & 6.45  & 6.62  & 6.62 \\
BP   & 2.12  & 2.02   & 1.65  & 2.27  & 2.05 \\
C    & 5.74  & 5.59   & 4.88  & 5.75  & 5.55 \\
MgO  & 9.16  & 9.06   & 8.34  & 9.52  & 9.22 \\
AlP  & 2.68  & 2.61   & 2.62  &       & \\
Si   & 1.31  & 1.21   & 0.96  & 1.29  & 1.29 \\
SiC  & 2.76  & 2.59   & 2.54  &       & \\
ZnO  & 4.54  & 4.46   &       &       \\
\bottomrule
\end{tabular}
\label{tab:eom_compare}
\end{table}

\begin{figure}[hbt!]
\centering
\includegraphics[width=0.47\textwidth]
{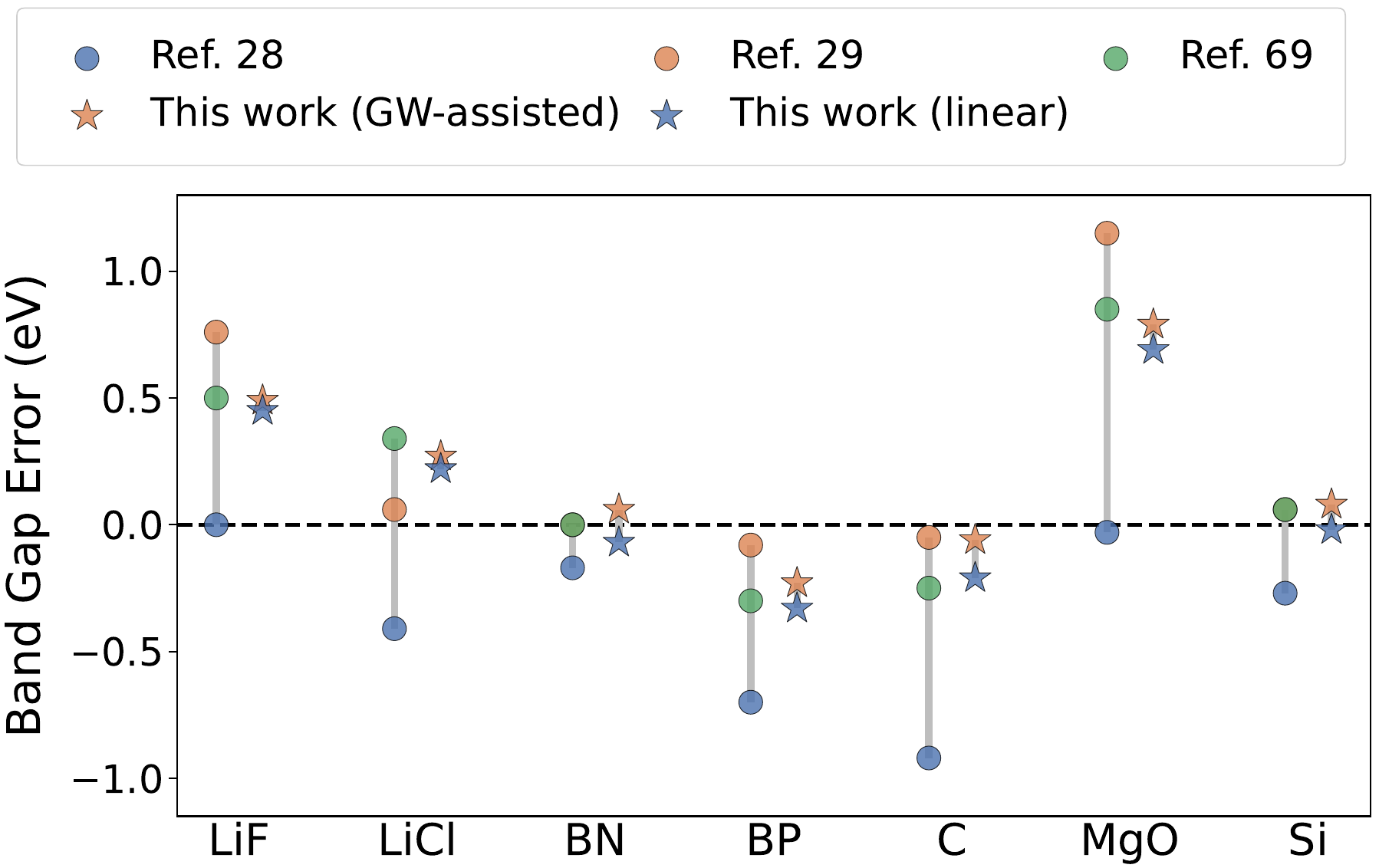}
\caption{Signed errors, $E_g^{\mathrm{calc}}-E_g^{\mathrm{exp,no\,ZPR}}$,
of thermodynamic-limit EOM-CCSD band gaps for the seven solids common
to this work and Refs.~\citenum{voPerformancePeriodicEOMCCSD2024}, \citenum{moermanExploringAccuracyEquationofmotion2025}, and \citenum{li2026reaching}. All errors are evaluated using the
experimental reference values listed in Table~II.}
\label{fig:compare_lterature}
\end{figure}

We summarize the basis-set-corrected TDL EOM-CCSD band gaps for all ten systems in Table~\ref{tab:eom_compare} and Fig.~\ref{fig:compare_lterature}, together with the corresponding literature values. 
Previous benchmark studies reported noticeably different TDL EOM-CCSD gaps for several systems. 
In Ref.~\citenum{voPerformancePeriodicEOMCCSD2024}, Vo \textit{et~al.} used a direct linear extrapolation based on the largest accessible \(3\times3\times3\) and \(4\times4\times4\) \(k\)-meshes, whereas in Ref.~\citenum{moermanExploringAccuracyEquationofmotion2025}, Moerman \textit{et~al.} used a \(GW\)-assisted extrapolation based on \(2\times2\times2\) to \(3\times3\times3\) \(k\)-meshes. 
For the seven systems common to both studies, the reported TDL gaps differ by as much as 1.18 eV, with particularly large discrepancies for LiF, BP, C, and MgO. These differences highlight the sensitivity of TDL EOM-CCSD extrapolations based on relatively small \(k\)-point meshes.

In the present work, the use of substantially denser \(k\)-meshes reduces this extrapolation sensitivity and mitigates the ambiguity associated with the choice of extrapolation scheme. Using \(5\times5\times5\) to \(8\times8\times8\) $k$-meshes, the \(GW\)-assisted and direct linear extrapolation schemes give very similar TDL gaps across the full test set: the two estimates differ by only 0.04--0.17 eV, with a mean absolute difference of approximately 0.10 eV. 
For MgO and Si, where calculations up to \(10\times10\times10\) are available, extending the extrapolation range from \(5\times5\times5\)--\(8\times8\times8\) to \(5\times5\times5\)--\(10\times10\times10\) changes the extrapolated gaps by only 0.03 and 0.02 eV, respectively, as shown in the SI. 
This stability indicates that the \(5\times5\times5\) and denser meshes are already close to the asymptotic finite-size regime for the systems considered here. Our \(GW\)-assisted results agree more closely with the \(GW\)-assisted extrapolations of Moerman \textit{et~al.}, with differences of up to 0.36 eV for the overlapping systems, which reflects the common use of \(G_0W_0\)@HF as an extrapolation coordinate that captures much of the leading finite-size behavior.

The new canonical periodic EOM-CCSD results of
Li \textit{et al.}~\cite{li2026reaching} provide another independent comparison
for the present calculations. Across the seven common solids (with same pseudopotential), the mean absolute
differences relative to Ref.~\citenum{li2026reaching} are only 0.07~eV for the GW-assisted
extrapolation and 0.08~eV for the direct linear extrapolation. This close agreement supports the reliability of both the
full-system and thermodynamic-limit extrapolations used in the present
work.

\subsection{Comparison of \texorpdfstring{$GW$}{GW} and EOM-CCSD}

\begin{table*}[t]
\centering
\begin{threeparttable}
\caption{
Thermodynamic-limit (TDL) band gaps of ten semiconductors and insulators obtained from $G_0W_0$@PBE, $G_0W_0$@HF, EOM-CCSD, and experiment. 
The $G_0W_0$ results were extrapolated using the cubic extrapolation scheme with \(k\)-meshes from \(3\times3\times3\) to \(8\times8\times8\).
The EOM-CCSD results were obtained from ibDET-derived full-system-limit gaps and extrapolated to the TDL using the \(GW\)-assisted scheme with \(k\)-meshes from \(5\times5\times5\) to \(8\times8\times8\). Basis set corrections to the quadruple-zeta level evaluated at $6\times6\times6$ $k$-mesh for $GW$ and $2\times2\times2$ $k$-mesh for EOM-CCSD were added to all calculated band gaps. 
Experimental gaps with zero-point renormalization (ZPR) removed are reported, with the observed experimental gaps shown in parentheses. ZPR corrections were taken from Ref.~\citenum{engelZeropointRenormalizationBand2022}, except for BP and LiCl, for which values from Ref.~\citenum{shangAssessmentMassFactor2021} were used. MSE/MAE/MARE stand for mean signed error, mean absolute error, and mean absolute relative error. 
All band gap values are in eV.}
\label{tab:summary}
\begin{ruledtabular}
\begin{tabular}{ccccc}
System & $G_0W_0$@PBE & $G_0W_0$@HF    & EOM-CCSD &  Exp.~no ZPR (obs.)   \\
\midrule
LiF      & 13.84 & 17.27 & 15.92 & 15.43 (14.2\cite{carotenutoThermoreflectanceWhiteTin1976}) \\
LiCl     & 8.96  & 11.46 & 10.11 & 9.84 (9.4\cite{Baldini1970})  \\
BN       & 6.36  & 8.59  & 6.68  & 6.62 (6.22\cite{raffertyDirectIndirectTransitions1998}) \\
BP       & 1.98  & 3.78  & 2.12  & 2.2, 2.5 (2.1\cite{raffertyDirectIndirectTransitions1998}, 2.4\cite{madelungSemiconductorsBasicData2012})\\
C        & 5.61  & 7.57  & 5.74  & 5.80 (5.48\cite{madelungSemiconductorsBasicData2012})  \\
MgO      & 7.57  & 10.83 & 9.16  & 8.37 (7.83\cite{whitedExcitonThermoreflectanceMgO1973})  \\
AlP      & 2.41  & 4.34  & 2.68  & 2.60 (2.51\cite{monemarFundamentalEnergyGaps1973})  \\
Si       & 1.11  & 2.86  & 1.31  & 1.23 (1.17\cite{madelungSemiconductorsBasicData2012}) \\
SiC      & 2.44  & 4.36  & 2.76  & 2.60 (2.42\cite{madelungSemiconductorsBasicData2012})  \\
ZnO & 2.52  & 6.19  & 4.54  & 3.52 (3.34\cite{kimStructuralReconstructionHexagonal2003,tekeExcitonicFineStructure2004})\\
\midrule
MSE & $-$0.56 & 1.89 & 0.27 & \\
MAE  & 0.56   & 1.89    & 0.32     &       \\
MARE & 10.3\%& 52.2\% & 7.2\%  &       \\
\end{tabular}
\end{ruledtabular}
\begin{tablenotes}
    \footnotesize
    \item[a] For BP, the midpoint of the reported experimental values was used in the error statistics.
    \item[b] For zinc blende ZnO, the reference gap was estimated from the zinc blende optical gap of 3.28 eV by adding the wurtzite ZnO exciton binding energy of 60 meV and ZPR correction of 175 meV.
\end{tablenotes}
\end{threeparttable}
\end{table*}

We next assess the accuracy of the TDL EOM-CCSD band gaps by comparing them with \(G_0W_0\) and experimental values with zero-point renormalization removed. 
Table~\ref{tab:summary} summarizes the band gaps of ten semiconductors and insulators obtained from \(G_0W_0\)@PBE, \(G_0W_0\)@HF, and EOM-CCSD using the same numerical framework in PySCF.
Basis-set corrections to the quadruple-zeta level were applied to all theoretical results. 
The \(G_0W_0\)@PBE results obtained here agree well with those of Ref.~\citenum{zhuAllElectronGaussianBasedG0W02021}, where \(k\)-meshes up to \(6\times6\times6\) and all-electron cc-pVTZ basis set were used, with typical differences of only about 0.1 eV.

We note that the observed and ZPR-corrected experimental gaps adopted
in Refs.~\citenum{moermanExploringAccuracyEquationofmotion2025} and \citenum{li2026reaching} are not entirely identical to those used here. The differences in the
observed gaps arise from the use of different experimental
literature sources. Direct determination of the fundamental gap,
particularly the conduction-band edge, is experimentally challenging,
and many reported values are inferred from optical absorption or
photoluminescence measurements together with an estimated exciton-binding
correction. The larger differences in the static-lattice reference gaps
additionally reflect the use of different calculated zero-point renormalization (ZPR) corrections,
which depend on details of the electron-phonon calculation. 

The comparison highlights the strong starting-point dependence of one-shot \(G_0W_0\) band gaps. 
The \(G_0W_0\)@PBE gaps are generally underestimated relative to experiment, giving a mean signed error (MSE) of $-$0.56 eV, a mean absolute error (MAE) of 0.56 eV, and a mean absolute relative error (MARE) of 10.3\%. 
This behavior is consistent with the overscreening error associated with semilocal starting points~\cite{gantOptimallyTunedStarting2022,liRenormalizedSinglesCorrelation2022}. 
In contrast, \(G_0W_0\)@HF substantially overestimates the band gaps for all systems, with an MSE/MAE of 1.89 eV and a MARE of 52.2\%, reflecting the underscreening associated with the HF starting point~\cite{gantOptimallyTunedStarting2022,liRenormalizedSinglesCorrelation2022}. 
Across the test set, the difference between \(G_0W_0\)@PBE and \(G_0W_0\)@HF ranges from 1.75 to 3.67 eV, demonstrating that the starting-point dependence of one-shot \(G_0W_0\) can be several electronvolts for these solids.

EOM-CCSD gives better overall accuracy than $G_0W_0$@PBE for the present test set, with an MSE of 0.27 eV, an MAE of 0.32 eV, and a MARE of 7.2\%. 
The EOM-CCSD errors are, however, system dependent. 
The absolute errors of EOM-CCSD are within 0.3 eV for most tested systems, while the largest deviations occur for MgO and ZnO, for which EOM-CCSD overestimates the reference gaps by 0.79 and 1.02 eV, respectively (a comparison on different basis set combinations for ZnO is included in the SI). 
These results indicate that EOM-CCSD provides competitive band gaps and captures the overall experimental trends across a chemically diverse set of solids, although residual system-dependent errors remain.

\subsection{Band structure of ZnO}
We further examine the performance of EOM-CCSD for describing the Zn \(3d\)-band position in ZnO. 
An accurate description of the Zn \(3d\) binding energy is important because the position of the \(3d\) manifold controls its hybridization with the O \(2p\) valence states and therefore affects the valence band structure and band gap. 
Experiment places the Zn \(3d\) band (in wurtzite ZnO) approximately 7.5 eV below the valence band maximum. 
However, DFT with conventional LDA/GGA functionals substantially underestimates this binding energy, typically by 2--3 eV, due to an incorrect description of the Zn-\(3d\)/O-\(2p\) hybridization. 

\begin{figure}[hbt!]
\centering
\includegraphics[width=0.47\textwidth]{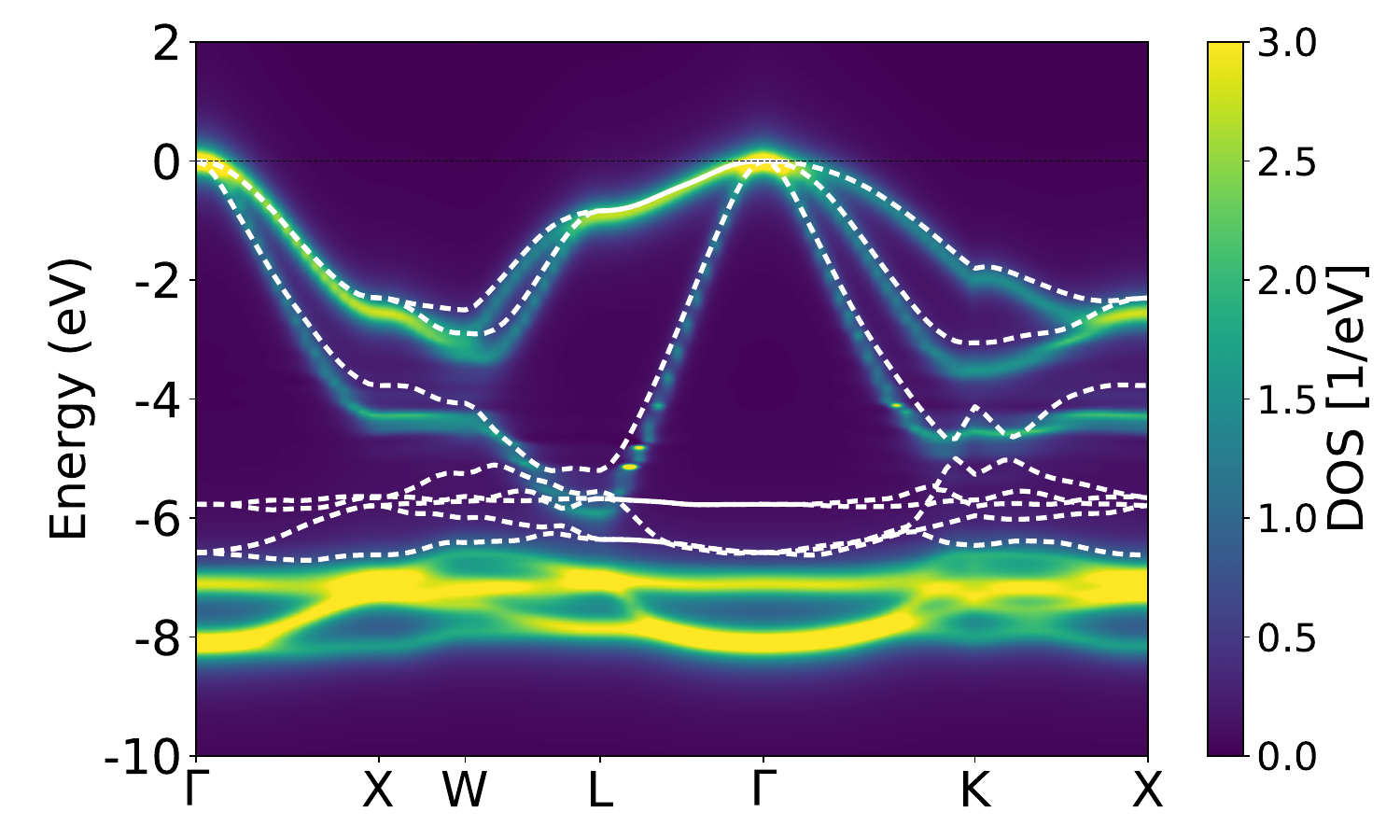}
\caption{Valence band structure of ZnO at the levels of EOM-CCSD (heat map) and $G_0W_0$@PBE (dashed line) with $8\times8\times8$ $k$-mesh. 
EOM-CCSD band structure was estimated by an ibDET calculation in which
$G_0W_0$@HF was used as the low-level theory and CCGF was used as the impurity solver.
The embedding space consists of 90 occupied orbitals and 270 virtual orbitals.
GTH-cc-pVDZ basis set and GTH-HF-rev pseudopotential were used for O.
The cc-pVDZ-PP and the corresponding effective core potential were used for Zn.}
\label{fig:zno_band}
\end{figure}

To compute the Zn \(3d\)-band position at the EOM-CCSD level, we apply ibDET using \(G_0W_0\)@HF as the low-level theory and CCGF as the impurity solver. 
The resulting ZnO band structures at the levels of EOM-CCSD and \(G_0W_0\)@PBE are shown in Fig.~\ref{fig:zno_band}. 
Following Ref.~\citenum{klimesPredictive$GW$Calculations2014}, we define the \(d\)-band position as the average energy of the Zn \(3d\) bands at the \(\Gamma\) point relative to the valence band maximum. We note that, although the present calculations are performed for zinc blende ZnO, recent DFT benchmark suggests that the Zn \(3d\)-band position is not sensitive to the ZnO crystal phase~\cite{bhattacharjee2024customizing}.
The \(G_0W_0\)@PBE calculation gives a Zn \(3d\) binding energy of 6.20 eV, in good agreement with previous \(GW\) results\cite{limAngleresolvedPhotoemissionQuasiparticle2012,klimesPredictive$GW$Calculations2014}, but it underestimates the experimental value by more than 1 eV.
In contrast, EOM-CCSD predicts a Zn \(3d\) binding energy of 7.60 eV, in close agreement with experiment. 
Thus, despite overestimating the band gap of ZnO by more than 1 eV, EOM-CCSD provides an accurate description of the \(d\)-band binding energy in ZnO. We thus suspect that the band gap overestimation by EOM-CCSD originates primarily from the conduction band quasiparticle energies, which may reflect an incomplete description of screening and orbital relaxation effects.

\FloatBarrier

\section{Conclusions}
In summary, we applied interacting-bath dynamical embedding theory (ibDET) to address finite-size errors in periodic EOM-CCSD band gap calculations of solids. By separately controlling the finite-embedding-size error and the finite-\(k\)-mesh error, we obtained EOM-CCSD band gaps for ten semiconductors and insulators using dense Brillouin-zone sampling, reaching up to \(10\times10\times10\) \(k\)-point meshes. Once \(5\times5\times5\) and denser meshes are used, the \(GW\)-assisted and direct linear finite-size extrapolation schemes yield closely consistent thermodynamic-limit gaps. This agreement demonstrates that dense \(k\)-point sampling substantially reduces the extrapolation ambiguity that has complicated previous periodic EOM-CCSD benchmarks.

Using these finite-size-controlled results, we carried out an equal-footing comparison of EOM-CCSD, \(G_0W_0\)@PBE, and \(G_0W_0\)@HF. 
For the present test set, EOM-CCSD gives a mean absolute error of 0.32 eV and a mean absolute relative error of 7.2\% relative to ZPR-removed experimental values. These errors are smaller than those of $G_0W_0$@PBE and substantially smaller than those of $G_0W_0$@HF, demonstrating the promise of EOM-CCSD for band gap predictions in solids. The remaining EOM-CCSD errors are system dependent, with the largest deviations occurring for MgO and ZnO. Although EOM-CCSD overestimates the ZnO fundamental gap by more than 1 eV, it predicts the Zn \(3d\) binding energy in close agreement with experiment and noticeably improves upon \(G_0W_0\)@PBE. Overall, this work demonstrates that ibDET offers a practical and systematically improvable route for bringing high-level wavefunction-based many-body methods to realistic periodic electronic-structure problems. 

\begin{acknowledgments}
This work was primarily supported by the National Science Foundation under Grant No.~CHE-2337991. 
The quantum embedding software infrastructure development was supported by the National Science Foundation under Grant No.~OAC-2513473. 
C.H. was partially supported by a fellowship from The Molecular Sciences Software Institute under NSF grant CHE-2136142. 
C.V. acknowledges support from the Department of Defense through the National Defense Science \& Engineering Graduate (NDSEG) Fellowship Program. 
We thank Shuhang Li and Timothy Berkelbach for sharing EOM-CCSD data and helpful discussions. 
We thank the Yale Center for Research Computing for guidance and use of the research computing infrastructure,
particularly Matt Petersen.
\end{acknowledgments}

\section*{Supporting Information}
Lattice constants and band gap positions,
band gaps obtained from $G_0W_0$@HF, $G_0W_0$@PBE, HF+$GW$, HF+CC with different $k$-meshes,
comparison between two extrapolation schemes,
basis set corrections for $G_0W_0$@HF, $G_0W_0$@PBE, and EOM-CCSD.

\subsection*{Conflict of Interest}
The authors have no conflicts to disclose.

\section*{Data Availability}
The data that support the findings of this study are available in the main text and supporting information.

\bibliographystyle{aipnum4-2}
\bibliography{ref}
\raggedbottom
\end{document}